# Enhanced Load Balancing Technique for SDN Controllers: A Multi-Threshold Approach with Migration of Switches

Mohammad Kazemiesfeh[a], Somaye Imanpour[b], Ahmadreza Montazerolghaem[c*]

[a]Faculty of Computer Engineering, University of Isfahan, Isfahan, Iran

[b]Faculty of Computer Engineering, University of Isfahan, Isfahan, Iran

[c]Faculty of Computer Engineering, University of Isfahan, Isfahan, Iran

*Corresponding author: Ahmadreza, Montazerolghaem; a.montazerolghaem@comp.ui.ac.ir

**Abstract**

Deploying multiple controllers in the control panel of software-defined networks increases scalability, availability, and performance, but it also brings challenges, such as controller overload. To address this, load-balancing techniques are employed in software-defined networks. Controller load balancing can be categorized into two main approaches: (1) single-level thresholds and (2) multi-level thresholds. However, previous studies have predominantly relied on single-level thresholds, which result in an imprecise classification of controllers or have assumed uniform controller capacities in multi-level threshold methods. This study explores controller load balancing with a focus on utilizing multi-level thresholds to accurately assess controller status. Switch migration operations are utilized to achieve load balancing, considering factors such as the degree of load imbalance of the target controller and migration efficiency. This includes evaluating the post-migration status of the target controller and the distance between the migrating switch and the target controller to select the appropriate target controller and migrating switch. The proposed scheme reduces controller response time, migration costs, communication overhead, and throughput rate. Results demonstrate that our scheme outperforms others regarding response time and overall performance.

## 1 Introductions

Software-defined networks facilitate network management by vertically dividing the network into three planes, including the data plane, the control plane, and the application plane [1,2]. The controller domain in the data plane contains several switches that are managed by that controller [3]. The communication between the controller and the data plane is through the user interface, the most common OpenFlow user interface. OpenFlow helps speed up controller management [4].

The control panel consists of multiple controllers arranged in a distributed manner [5], each interconnected to monitor the network's status. However, these controllers can themselves become overloaded [6]. The need for multi-controller load balancing in SDN environments arises from the growing demand to manage an increasing number of devices and network traffic. Traditional network architectures, which rely on distributed control planes, may struggle to scale effectively and handle these rising demands. Multi-controller load balancing addresses this challenge by distributing the control plane workload across multiple controllers, thereby enhancing network scalability, availability, and the ability to manage a larger volume of devices and traffic [7]. One approach to mitigating controller load imbalance in Software-Defined Networking (SDN) environments is switch migration [8]. This method involves transferring a switch from the domain of an overloaded controller to that of an underloaded controller, thereby optimizing load distribution and improving overall network performance.

Despite the potential of switch migration operations and multi-controller frameworks, several gaps remain in the existing research. First, many approaches rely on simplistic single-level thresholds or assume uniform controller capacities, leading to inaccuracies in dynamic environments. Second, the high cost of switch migration often remains unaddressed, resulting in suboptimal network performance. Finally, current load evaluation methods frequently overlook the importance of using multiple metrics, such as CPU usage, memory consumption, and bandwidth, to comprehensively assess controller loads. These gaps highlight the need for a more accurate, efficient, and practical load-balancing framework for software-defined networks.

Recent advancements in software-defined networks (SDNs) and their extension to satellite networks highlight the growing complexity and potential of these systems. The integration of multi-domain networks in the 6G era, aimed at supporting applications like extended reality (XR) and holographic communications, faces challenges such as domain isolation and resource inefficiencies [9]. Similarly, studies in software-defined satellite networks emphasize the importance of optimizing multi-controller placement using meta-heuristic approaches, such as the Adaptive Virtual Bee Optimization Algorithm (AVOA), to enhance network reliability and control delay [10]. Furthermore, techniques like Virtual Network Request Load Balancing Profit (VNR_LBP) demonstrate the ability to mitigate congestion by virtualizing switches and optimizing resource utilization, leading to notable improvements in throughput, delay, and cost efficiency [11]. These recent developments underscore the necessity of robust frameworks to address the dynamic requirements of modern SDNs and their satellite counterparts.

This article focuses on load balancing among controllers in software-defined networks, aiming to accurately determine controller statuses using multiple thresholds while accounting for their varying capacities. To achieve efficient load distribution, switch migration operations are employed, strategically grouping controllers at different levels to minimize unnecessary migrations. Given the high cost associated with switch migrations, the approach prioritizes reducing migration overhead by selecting the most suitable switch and target controller with optimal efficiency. The proposed framework's performance is evaluated based on response time criteria through the design and implementation of various scenarios.

In this regard, the main contributions of this study are as follows:

1. Development of a Multi-level Approach for Determining Controller Status: This paper introduces an innovative method for accurately determining the load status of controllers using multiple thresholds. This approach significantly enhances the precision in identifying overloaded and underloaded controllers.

2. Load Calculation of Controllers Using Multiple Metrics: We propose a method for calculating the load on controllers that incorporates three key metrics: CPU usage, memory consumption, and bandwidth utilization. This approach allows for a more precise assessment of the load status of controllers and facilitates optimal load distribution.

3. Optimized Switch Migration Operations: This research investigates and optimizes the switch migration operations. By employing novel techniques, switches selected for migration are strategically chosen to reduce the load on overloaded controllers, thus improving the network's overall efficiency.

The structure of the paper is outlined as follows: Section 2 covers related works, while Section 3 elaborates on the proposed scheme. Section 4 details the evaluation and implementation setup and discusses the performance results. Finally, Section 5 presents the conclusions of the study.

## 2 Related works

In this section, we introduce the related works that have addressed the load balancing of controllers. Load balancing in software-defined networks has been widely studied under two main approaches: single-threshold and multi-threshold methods. Single-threshold methods define a specific threshold to classify controllers as overloaded or underloaded, while multi-threshold methods use multiple levels to assess controller states more dynamically and accurately.

### 2.1 Single Threshold

Reference[13] suggests a software-defined multiple load balancing strategy that relies on response time. This approach first measures the response time of the controllers. The controller's response time determines the threshold and is divided into overload and underload controllers. Reference[14], Xiang et al. proposed a deep reinforcement learning architecture for software-defined networks to solve the switch migration problem. In this paper, a model is designed for DDQN, which decides how to transfer switches between controllers. Reference[15] introduces an efficient switch migration-based load balancing (ESMLB) framework. This framework explores the use of software-defined networks (SDN) with multiple controllers to enhance reliability and scalability in Edge Computing (EC) environments, particularly in the context of the Internet of Things (IoT). The proposed ESMLB framework focuses on efficient switch migration for load balancing, addressing issues such as uneven load distribution among controllers and suboptimal performance caused by dynamic network demands and changing topologies. By using a decision analysis method to rank controllers based on criteria such as memory usage and CPU load, ESMLB seeks to optimize resource utilization and improve overall performance in SDN-based IoT systems.

Zafar et al.[16] have introduced a dynamic switch migration-based load balancing (DSMLB) approach to tackle the challenges associated with managing Internet of Things (IoT) traffic in Software-Defined IoT (SD-IoT) networks. This framework proposes a dynamic switch migration strategy to optimize load balancing performance, addressing the limitations of existing techniques in handling real-time traffic variations and diverse network topologies. By implementing the DSMLB framework, the authors aim to enhance load balancing by dynamically transferring switches from overloaded to underutilized controllers, considering key metrics like controller input overhead, control plane request rate, and capacity. Through experimental testing on the Mininet platform with the RYU controller, the DSMLB mechanism shows significant enhancements in reducing controller response time, and migration cost, and improving control plane load balancing efficiency compared to standard methods, underscoring its effectiveness in managing controller workload within SD-IoT architectures. The authors of [17] address the issue of uneven load distribution among distributed controllers in Software-Defined Networking (SDN) by introducing a dynamic and adaptive load balancing mechanism. The mechanism utilizes a hierarchical control plane to estimate the load of each controller based on controller and switch factors. The article[18] addresses the issue of load imbalance in the software-defined networking (SDN) control plane by proposing switch migration as a solution. This involves moving switches from overloaded controllers to underutilized ones to improve network performance and resource utilization.

Reference[19] presents GLBMF, a load-balancing method driven by greediness that focuses on reducing switch migrations and giving priority to low-traffic mice flows. GLBMF operates by redistributing switches from heavily loaded controllers to lightly loaded ones when the deviation from the average load surpasses a predefined threshold. Moreover, it gives precedence to switches with minimal mice-flow traffic for relocation to minimize network-wide packet losses. Experimental results indicate that GLBMF surpasses conventional methods in terms of throughput, switch movements, response time, and packet loss mitigation. In the paper [20], a holonic multi-agent-based approach is introduced to solve the controller placement problem in SDN-SG networks. Using this method, the time complexity of the problem is significantly reduced, and load balancing is effectively distributed across the network.

## 2.2 Multi Threshold

Reference[21], they have proposed the migration switch scheme of Multi Threshold Load Balancing (MTLB). This design first sets the multi-level threshold and then determines the status of the controllers, if the controller level changes, the rest of the controllers are notified, and finally, by carefully studying the behavior of the migrant switch and the target controller, it has provided a method for the migration operation.

Based on the articles that have been explained and a comparison has been made between them in Table I, most of the articles in the field of controller load balancing have problems of a single domain, high switch migration cost, lack of selection of the appropriate target controller, and computational complexity. First, we deal with the problem of a single province. Single states do not provide accurate classification of controllers, leading to inefficient load balancing. However, the paper [23] adopted a multi-threshold approach by considering the number of incoming packets to the controller, which does not provide a precise criterion for evaluating the controller load. However, in this article, the problem is that all controllers have the same capacity. Therefore, to solve this problem, in this article, an attempt has been made to use controllers with different capacities and combine them with the multi-level threshold approach to provide accurate scaling and classification of controllers. In addition, in this plan, for accurate measurement of the controller load, CPU consumption criteria, memory usage, and bandwidth are used to evaluate the controller load. Then, by an algorithm that does not have the complexity of calculations; the target switch and the appropriate target controller are selected to solve the problem of choosing the proper target controller and the problem of high switch migration costs.

Table 1: The Summary of the Notations that are used

| References | Parameters | Threshold | controller | Advantage | Dis-advantage |
|---|---|---|---|---|---|
| [13] | Response time | Single | Floodlight | Quickly balancing the load of controllers | High switch migration cost |
| [14] | Packet-in | Single | Ryu | Enhancing load balancing | High time overhead |
| [15] | CPU, Memory, Bandwidth | Single | Floodlight | reducing migration times | Not suitable for small environments |
| [16] | CPU, Memory, Bandwidth | Single | Ryu | Reducing response time | High cost and complexity |
| [18] | Packet-in | Single | DHA-CON | Adequate use of controller resources | Static connection between switch and controller |
| [19] | Packet-in | Single | Ryu | Increase network throughput, reduce response time | Equal consideration of the capacity of all controllers |
| [21] | Packet-in | Multiple | Floodlight | Controller overhead is low | Equal consideration of the capacity of all controllers |

## 3 Proposed framework

In this section, we will describe the proposed plan. The CPU, memory, and bandwidth criteria are used to calculate the controller load in this plan. Next, the controller load is compared against the multi-level thresholds to determine the controller's level. Based on the load level of the controller, which is at a high load level or overloaded, the switch migration operation is performed.

### 3.1 System Model

As you can see in Fig. 1, the system model of the proposed design includes several controllers in the control plane, where each controller in the data plane contains a domain. The domain of each controller also includes several switches, the number of switches in the domain of each controller is based on the capacity of that controller. The controllers are also connected so that each controller is independently aware of the status of the entire network.

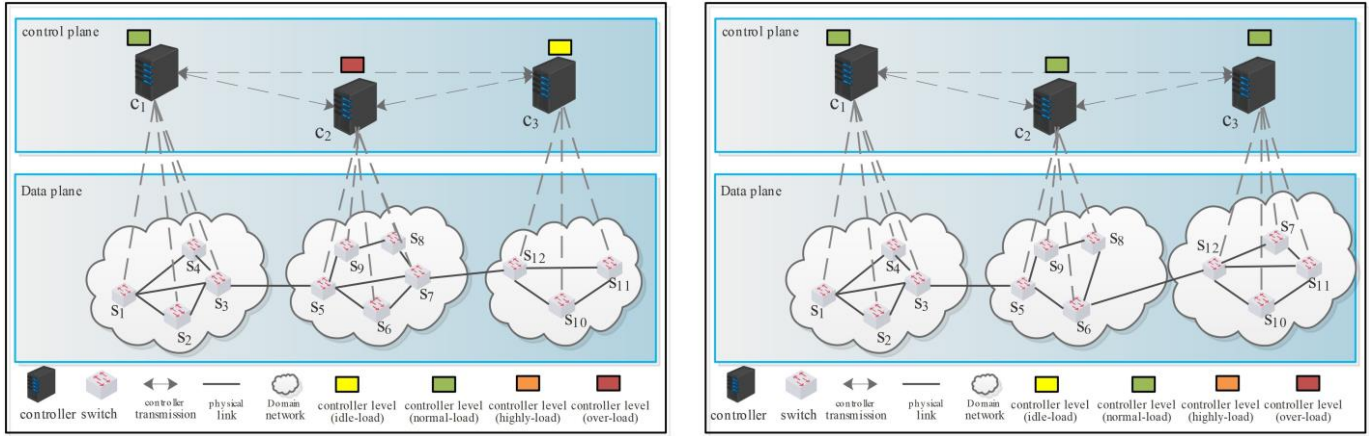

Figure 1: System model

The network consists of switches and controllers. Let $C = \{c_1, c_2, \ldots, c_j, \ldots, c_K\}$ represent the set of controllers, where $K$ denotes the total number of controllers in the network. Similarly, $S = \{s_1, s_2, \ldots, s_i, \ldots, s_N\}$ represents the set of switches, with $N$ as the total number of switches in the network. As mentioned earlier, each controller manages a specific subset of switches. Consequently, the total number of switches within the domain of controller $c_j$ is denoted as $N'$. The notation $s^j_i$ refers to a switch located within the domain of controller $c_j$. As illustrated in Figure 1, which depicts the flowchart of the prediction algorithm, the process begins with measuring the controller's load. In the second step, a multi-level threshold mechanism is applied to classify controllers into different levels. The third step involves the load balancing operation, which, if necessary, includes selecting an appropriate target switch, identifying a suitable target controller, and finally, executing the switch migration process.

## 3.2 Framework flowchart

The operational flow of the proposed DLBMT framework is illustrated in Fig. 2, which outlines the step-by-step load balancing process, from load monitoring to switch migration.

1. load measurement of the controller: the controller load is calculated based on CPU usage, memory, and bandwidth metrics. The controller load is a numerical value between 0 and 100.

2. multilevel threshold: Four thresholds are available for determining the controller level: idle level, normal level, high load level, and overload level. Subsequently, the controller's level is determined based on its load.

3. Notify controller: If the controller's level changes, other controllers are informed.

4. Status of the controller: When a controller is at the normal level, it can manage the domain effectively. However, when a controller is in the idle, high load, or overload level, a switch migration operation is executed if a suitable migration switch and target controller pair are available.

5. ratio of the switch's consumed resources to the distance of the switch from the controller: For load balancing operations, the ratio of the resources consumed by the source controller to the distance between the switch and the source controller is calculated.

6. Selecting appropriate migration switches from the available switches within the domain of the source controller: The calculated value is compared for all switches within the domain to the average resources consumed by the source controller to the distance between the switch and the source controller. Switches that have a higher resource consumption ratio compared to the average for the source controller are added to the migration switch set.

7. Determining the target controller's load status (whether it is operating at a normal or idle level): non-source controllers in the idle or normal level are selected as target controllers.

8. Determining the target controller's load status (whether it is operating at a normal or idle level): The load levels for the target controllers after switch migration are calculated for each switch in the migration switch set.

9. Determining the source controller's load status after the switch migration.

10. Removing the controller from the list of suitable target controllers of the switch: If a target controller is not in the idle or normal level after switch migration, it is removed from the list of target controllers for the switch in the migration switch set.

11. Calculating the degree of load imbalance between the source controller and the target controller: For the remaining target controllers, the degree of load imbalance between the source controller and the target controller is calculated.

12. Choose the most appropriate target controller for each switch and save it: For the remaining target controllers, the degree of load imbalance between the source controller and the target controller is calculated. Among the target controllers in the migration switch set, the target controller with the least load imbalance degree is selected as the suitable target controller for the switch. This pair of the target controller and switch is stored in a set.

13. Any pair with a lower efficiency is selected for the migration operation: If the migration switch set and target controller set are not empty, the migration efficiency is calculated for all available pairs in the set. The target controller and migration switch with the least migration efficiency are selected for the switch migration operation.

14. turn on a controller or add a controller to the network or do not have switch migration: If the migration switch set and target controller set are empty, and the source controller is idle or at a high load level, no switch migration operation is performed. However, if the source controller is at the overload level and there is a previously turned-off controller available, it is turned on, and the load-balancing operation is carried out. Otherwise, a new controller is added to the network.

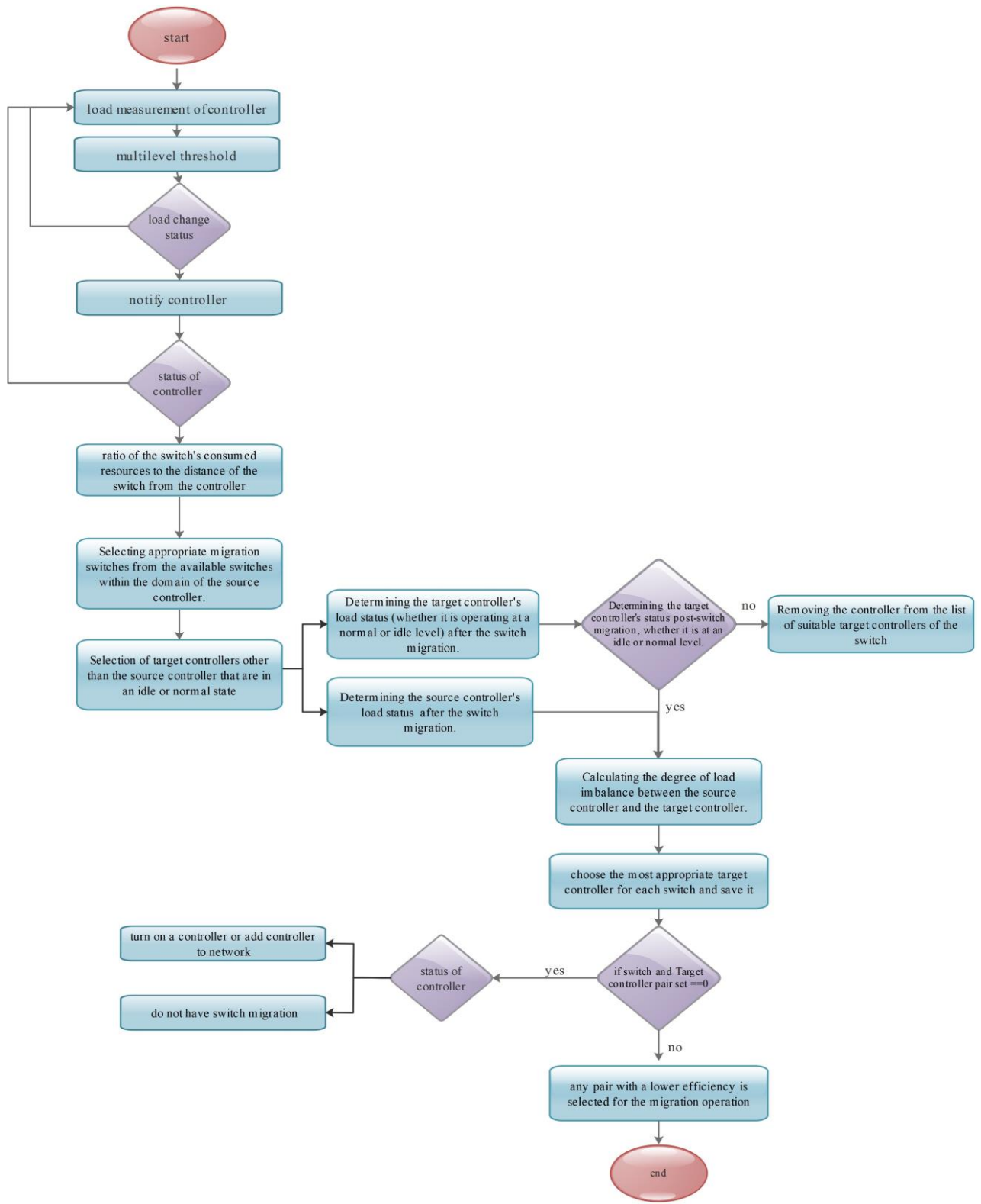

Figure 2: Framework flowchart

All the symbols used in this article are listed in Table 2.

Table 2: The Summary of the Notations that are used.

| Symbol | Description |
|---|---|
| $c_j$ | controller |
| $s_i^j$ | The switch in the controller domain $c_j$ |
| $\pounds_i^j$ | Total use of controller resources by a switch $s_i^j$ in the controller domain $c_j$ |
| $\psi_{s_i^j}$ | The ratio of resources consumed by the switch to the distance between the switch and the controller |
| $h_{mi}$ | Distance switch $s_i^j$ to controller $c_j$ |
| $LR_{c_j}$ | Load controller $c_j$ |
| $LR_{c_k}$ | Load controller $c_k$ |
| $LR_{c_j}^*$ | The amount of load of the source controller $c_j$ after the migration of the switch $s_i^j$ |
| $\pounds_{ij}^k$ | The consumption rate of the switch muhajirs $s_i^j$ from the resources of the target controller $c_k$ |
| $DC_{(c_j,c_k)}^*$ | Degree of load imbalance between source controller $c_j$ and target controller $c_k$ for switch $s_i^j$ |
| $LR^*$ | Average load of controllers after migration |
| $\theta_{j,k}$ | Migration efficiency |
| $f_{(s_i^j, c_k)}$ | The cost of immigration |
| $h_{ik}$ | Distance switch $s_i^j$ to target controller $c_k$ |

## 3.3 Formulation proposed algorithm

### 3.3.1 Load Measurement

To ensure efficient network management, it is crucial to accurately measure the load on each controller. The load measurement process involves evaluating the resource consumption of the controllers based on interactions with their associated switches. This section presents the proposed method for calculating controller load, focusing on CPU, memory, and bandwidth usage. In the first step, the controller load is calculated using the proposed design. To calculate the load of the controller, first, the consumption of CPU, memory, and bandwidth of each controller is calculated by the packet-in requests of the switches in the domain of that controller, which is as follows:

$$\pounds_i^j = a \cdot \left(\frac{Load_{CPU}}{CPU}\right) + b \cdot \left(\frac{Load_{Mem}}{Mem}\right) + c \cdot \left(\frac{Load_{Bw}}{Bw}\right) \tag{1}$$

$$\pounds^j = \sum_{s_i^j \in c_j} \pounds_i^j \tag{2}$$

$$LR_{c_j} = \pounds^j * 100 \tag{3}$$

Equation (1): $\pounds_i^j$ represents the total resource usage of controller $c_j$ by switch $s_i^j$ within its domain. $Load_{CPU}$, $Load_{Mem}$, and $Load_{Bw}$ denote the respective consumption of CPU, memory, and bandwidth by switch $s_i^j$ on controller $c_j$. Meanwhile, $CPU$, $Mem$, and $Bw$ represent the total capacities of CPU, memory, and bandwidth for the controllers. The coefficients $a$, $b$, and $c$ are used in the calculation, with their sum equaling one. Equation (2) calculates the total resources consumed by the switches connected to controller $c_j$. Equation (3) then converts this total into a percentage, yielding a value between one and one hundred. This calculated value for each controller $c_j$ represents the load of the controller.

### 3.3.2 Multi-level threshold

After calculating the controller load, the controller's level is determined. As mentioned earlier, the controller load is a value between 0 and 100. As illustrated in Fig. 2, the value of 100 is divided into four levels, with three thresholds: the first threshold is 25, the second is 50, and the third is 75. If the controller's load is less than the first threshold (i.e., between 0 and 25), it is classified as idle. If the load is between 25 and 50, the controller is considered to be at a normal level. If the load is between the second and third thresholds (i.e., between 50 and 75), the controller is classified as having a high load. At this level, if conditions permit, a switch migration operation is performed. Finally, if the controller's load exceeds the third threshold (above 75), it is classified as overloaded, and a switch migration operation must be executed. This operation is carried out even if there is no target controller for migration; in such cases, a new controller is added to the network.

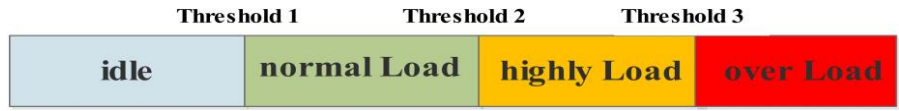

Figure 3: Multi-level threshold

In Algorithm 1, the process of determining the controller level has been performed. In this algorithm, the controller load, the current level of the controller (denoted by $q_j$), and the list of levels are taken as inputs. The list of levels, Q[i], includes four values: Q[1]=25, Q[2]=50, Q[3]=75, and finally Q[4]=100. The algorithm's output is a new level for the controller and a boolean value that indicates whether the level of the controller has changed or not.

In line 3 of the algorithm, a loop is created to consider all the controllers in the network. In line 4, there is an internal loop for the levels, with a maximum value of 4. In lines 5 and 6, the new controller load variable is set to zero, and the current load of the controller is also checked. In line 7, there is an IF condition for comparing the controller load with the levels to determine the level of the controller.

Based on the example network depicted in Figure 1, which consists of three controllers, each with multiple switches in its domain: in the domain of controller $c_1$, there are four switches $\langle s_1^1, s_2^1, s_3^1, s_4^1 \rangle$; in the domain of controller $c_2$, there are three switches $\langle s_1^2, s_2^2, s_3^2 \rangle$; and in the domain of controller $c_3$, there are two switches $\langle s_1^3, s_2^3 \rangle$. The controller loads are $LR_{c_1} = 76$, $LR_{c_2} = 47$, and $LR_{c_3} = 21$.

The level determination algorithm is applied to all three controllers, resulting in the following classifications: controller $c_1$ is in the overload level, controller $c_2$ is at the normal level, and controller $c_3$ is in the idle level. To determine the level for controller $c_2$ with a load of 47, the controller's load is first compared with Q[1] (which is 25). Since the condition is met, the variable $i$ is incremented by 1, and the load is then compared with Q[2] (which is 50). The condition holds again because 47 < 50. As a result, in line 6, the value of $i$ (which is 2) is stored in the $c_j$ variable, and the loop terminates.

When the value of $c_j$ is equal to 1, it indicates that the controller is in the idle state. If $c_j = 2$, the controller is at a normal level. If $c_j = 3$, the controller is at a high-load level, and finally, if $c_j = 4$, the controller is in an overload state. In line 11, the new controller level is compared with its current level. If the level has changed, the algorithm outputs a value of True; otherwise, it outputs False.

---

**Algorithm 1** Multi-level threshold

1: **Input:** $LR*_{c_j}$ current controller load level: $q_j$ , list of thresholds: Q[i]
2: **Output:** $c_j$ , T= true or F=false
3: **for** $j = 1$ **do**
4:     **for** $i = 1$ **do**
5:         $c_i \leftarrow 0$
6:         **Check** $q_j$
7:         **if** $LR_{c_j} < Q[i]$ **then**
8:             $c_j \leftarrow i$
9:             **break**
10:        **end if**
11:        $i \leftarrow i + 1$
12:     **end for**
13:     **if** $q_j = c_j$ **then**
14:         **return** false
15:     **else**
16:         **return** $q_j = c_j$
17:         **return** true
18:     **end if**
19:     $j \leftarrow j + 1$
20: **end for**

### 3.3.3 Load balancing

This section discusses the controller load balancing operation, with the goal of maintaining the controller load at a normal level. Once the levels of the controllers are assessed, any changes in a controller's load level will trigger notifications to the other controllers. If a controller in the network is operating at an under-load level, we will attempt to deactivate it based on the network conditions.

If, in addition to the source controller operating at an under-load level, there is another controller in the network that is either at an under-load or normal load level, and this controller continues to operate at an under-load or normal level after the switch migration operation, the migration will be executed, and the source controller will be deactivated.

When a controller operates at a normal load level, it indicates that the controller can effectively manage its domain, and efforts will be made to sustain the controllers' loads in the network at this level. In contrast, if a controller reaches a high load level, it signifies that the controller is approaching overload, prompting the execution of load balancing operations for controllers at this high load threshold.

To perform load balancing operations for controllers at a high load level, we first calculate, for all switches in the domain of the high-load controller, the ratio of total resource usage by switch $s_i^j$ in the domain of controller $c_j$ to the distance of switch $s_i^j$ from the controller $c_j$. This is done using the following equation (4):

$$\psi_{s_i^j} = \frac{\pounds_i^j * 100}{h_{mi}} \tag{4}$$

In this equation, $\psi_{s_i^j}$ represents the ratio of switch resource consumption to the distance of the switch from the controller, while $\pounds_i^j$ denotes the total resource consumption of controller $c_j$ by switch $s_i^j$ in its domain. $h_{mi}$ represents the distance of switch $s_i^j$ from controller $c_j$. We then calculate the average value of this ratio for all switches in the domain of controller $c_j$ and store the value of $\psi_{s_i^j}$ for each switch in an array. If the value exceeds the average, the array containing switches selected from the source controller's domain is referred to as the migration switch array.

$$LR_{c_j}^* = [LR_{c_j} - (\pounds_i^j * 100)] \tag{5}$$
$$LR_{c_k}^* = [LR_{c_k} + (\pounds_{ij}^k * 100)] \tag{6}$$

In Equation (5), the value $LR^*c_j$ represents the load of the source controller $c_j$ after the migration of the selected switch $s_i^j$ from the migration switch array. In this equation, $LR_{c_j}$, which denotes the load of the source controller $c_j$, is subtracted by $\pounds_i^j$, which indicates the total resource consumption by the switch $s_i^j$ on the controller $c_j$. Since $\pounds_i^j$ is a value between 0 and 1, it is multiplied by 100 in this formula to yield a value between 0 and 100. In Equation (6), $LR^*{c_k}$ represents the load of the target controller $c_k$ after the migration of the switch. The term $\pounds_{ij}^k$ indicates the resource consumption of the selected switch $s_i^j$ from the migration switch array on the target controller. In this equation, $LR_{c_k}$, which denotes the load of the target controller $c_k$, is added to $\pounds_{ij}^k$, representing the total resource consumption of controller $c_k$ by switch $s_i^j$. The value of $\pounds_{ij}^k$ is calculated as follows:

$$\pounds_{ij}^k = a \cdot \left(\frac{Load_{CPU}}{CPU_k}\right) + b \cdot \left(\frac{Load_{Mem}}{Mem_k}\right) + c \cdot \left(\frac{Load_{Bw}}{Bw_k}\right) \tag{7}$$

In equation (7), the value of $CPU_k$, $Mem_k$ and $Bw_k$ indicates the maximum capacity of the target controller $c_k$.

$$DC^*_{(c_j,c_k)} = \frac{\frac{1}{2}\left(\sqrt{(LR_{c_j}^* - LR^*)^2} + \sqrt{(LR_{c_k}^* - LR^*)^2}\right)}{LR^*} \tag{8}$$

Equation (8), the degree of load imbalance between the source controller $c_j$ and the target controller $c_k$ is calculated for the switch $s_i^j$, and $LR^*$ represents the average load of the controllers after migration.

We first identify and pair each controller that remains idle or in a normal state after the migration of switch $s_i^j$ and store these pairs in a separate array. Since multiple target controllers may be available for a given switch, we evaluate all possible pairs and select the controller with the lowest degree of load imbalance. In other words, we choose the most suitable target controller for each switch and save the selected pair. If no valid switch-controller pairs are found, the process stops. Otherwise, the migration efficiency is calculated for all pairs, and the one with the highest efficiency is chosen for the migration operation. The efficiency calculation is as follows:

$$\theta_{j,k} = \frac{|DC^*_{(c_j,c_k)} - DC_{(c_j,c_k)}|}{f_{(s_i^j,c_k)}} \tag{9}$$

$$f_{(s_i^j,c_k)} = (\pounds_{ij}^k * 100) \cdot min(h_i k) \tag{10}$$

As you can see, the migration efficiency is calculated in equation (9), which is calculated by deducting the difference in the degree of load imbalance between the source controller $c_j$ and the target controller $c_k$ after the switch migration operation and the load imbalance degree between the source controller $c_j$ and the target controller ck before the operation. Switch migration is calculated and divided by migration cost. In this formula, $\theta_{j,k}$ represents migration efficiency and $DC^*_{(c_j,c_k)}$ represents the degree of load imbalance between two controllers before the migration operation, and $f_{(s_i^j, c_k)}$ represents Pays for immigration. Equation (10), the cost of migration is calculated by multiplying the amount of resources consumed by the migrator switch $s_i^j$ from the target controller ck by the distance between the migrator switch $s_i^j$ and the target controller ck. In this formula, $h_{ik}$ represents the distance of switches $s_i^j$ to the target controller ck. Finally, each migrating switch and target controller that has the lowest efficiency is selected for the switch migration operation.

For the controllers that are overloaded, we also have the switch migration operation. All the steps to perform the switch migration operation are the same as for the controllers that are overloaded, with the difference that if the switch array and the appropriate target controller were empty, if we had a switched-off controller in the network We turn it on and perform the migration operation. Otherwise, we add a controller to the network and perform the switch migration operation.

Algorithm 2 for load balancing between controllers using switch migration is for controllers that are in idle, high, or overload levels. Controllers at the normal level can manage their domains well. In steps 2-9, switches in the domain of the source controller that are suitable for migration are stored. In step 2, there is a loop for all switches in the domain of the source controller. In step 3, the ratio of the resources consumed by the switch to the distance from the switch to the source controller is calculated using formula 4. Then, in step 5, the total resources consumed by the switch from the source controller are calculated. After performing calculations for all switches in the domain of the source controller, the average resources consumed by the switch from the source controller are calculated in step 6. Then, using the condition in step 7, any switch that has a value greater than the average is stored in the P array.

In the given example, controller $c_1$ is in an overload state, and to achieve balance in the network, load-balancing operations must be carried out for controller $c_1$. Initially, the resource consumption of all switches in the domain of controller $c_1$ is calculated, yielding the following values: $< 20, 14, 7, 6 >$. Next, the values of $\psi_{sj}$ are computed: $\psi_{s_1^j} = 47$ and $\psi_{s_2^j} = 12$. Since switches $s_1^1$ and $s_2^1$ have higher resource consumption than the average, they are stored in the migration array, denoted as $\rho$.

After selecting the appropriate switches for migration, the next step is to identify a suitable destination controller. The primary condition for a destination controller is that it must be in either an idle or normal load state. Following this, the resource consumption of both the source controller and the destination controller after migration is calculated. The load level of the destination controller after migration is then determined. If the destination controller falls into a high-load or overload state after migration, it is removed from the list of potential target controllers. Conversely, if the destination controller remains in an idle or normal load state, the degree of imbalance in resource consumption between the source and destination controllers is calculated.

In this example, switches $s_1^1$ and $s_2^1$ are selected for migration. Controllers $c_2$ and $c_3$ are considered potential target controllers. First, the resource consumption of the source controller and potential destination controllers after the migration is calculated for switch $s_1^1$. Specifically, $LR_{c_1} = 56$, $LR_{c_2} = 78$, and $LR_{c_3}^c = 46$. Since controller $c_2$ enters the overload level after migration, switch $s_1^1$ is removed from the list of potential destination controllers. Therefore, the only remaining viable pair is switch $s_1^1$ and controller $c_3$. The same process is repeated for switch $s_2^1$, with the following values: $LR_{c_2} = 47$, $LR_{c_1} = 90$, and $LR_{c_3}^c = 55$. In this case, both controllers $c_2$ and $c_3$ enter the overload level after migration, meaning neither of them can be considered as a potential destination controller.

Finally, to select the best pair of switches and controllers for migration, the pairs with the highest migration rate are selected according to steps 30-35. If there are multiple pairs of switches and controllers for migration, the pair with the highest migration rate is selected. In this example, the pair ($s_2^1$ $C_3$) with a migration rate of 0.009 is selected. After selecting the pair of switches and controller for migration and saving them in Y, steps 25-29 of the algorithm stop the migration operation if =0 and the source controller is in the idle or normal level, as a suitable destination controller has not been found. However, if the source controller is in the high or overload level, the previously turned-off controllers are turned on again or the network is expanded and additional controllers are added to the network.

**Algorithm 2** Load Balancing

1: **Input:** $c_j$, $s_i^j$, $h_{mi}$, m, $LR_{c_j}$, $LR_{c_k}$
2: **Output:** SC
3: **if** $c_j == 1 || c_j == 3 || c_j == 4$ **then**
4:     **for** $s_i^j \in c_j$ **do**
5:         $\psi_{s_i^j} = \pounds_i^j * 100/h_{mi}$
6:         $\psi_{sj} = \psi_{sj} + \psi_{s_i^j}$
7:     **end for**
8:     $\overline{\psi_{sj}} = \frac{1}{m}\psi_{sj}$
9:     **if** $\psi_{s_i^j} >= \overline{\psi_{sj}}$ **then**
10:         $\rho \leftarrow \psi_{s_i^j}$
11:     **end if**
12:     **for** $\psi_{s_i^j} \in \rho$ **do**
13:         $LR^*_{c_j} = [LR_{c_j} - (\pounds_i^j * 100)]$
14:         $LR^*_{c_k} = [LR_{c_k} - (\pounds_{ij}^k * 100)]$
15:         $\pounds_{ij}^k = a \cdot (\frac{Load_{CPU}}{CPU_k}) + b \cdot (\frac{Load_{Mem}}{Mem_k}) + c \cdot (\frac{Load_{Bw}}{Bw_k})$
16:         **The process of determining the level according to Algorithm 1 for** $LR^*_{c_k}$ **is carried out, and the output is stored in** $C^*_k$
17:         $DC^*_{(c_j,c_k)} = \frac{\frac{1}{2}(\sqrt{(LR^*_{c_j}-LR^*)^2} + \sqrt{(LR^*_{c_k}-LR^*)^2})}{LR^*}$
18:     **end for**
19:     **while** $\rho > 0$ **do**
20:         **for** $C^*_k == 1 || C^*_k == 2$ **do**
21:             **if** $DC^*_{(c_j,c_k)} < tmp$ **then**
22:                 $tmp \leftarrow DC^*_{(c_j,c_k)}$
23:             **end if**
24:             $\gamma(s_i^j, c_k)$
25:         **end for**
26:     **end while**
27:     **if** $\gamma == 0$ **then**
28:         **if** $c_j == 1 || c_j == 3$ **then**
29:             **The switch migration operation is not carried out.**
30:         **if** $c_j == 4$ **then**
31:             **A controller is added to the network.**
32:         **else**
33:             **for** $(s_i^j, c_k)$ **do**
34:                 $\theta_{j,k} = \frac{|DC^*_{(c_j,c_k)} - DC_{(c_j,c_k)}|}{f_{(s_i^j,c_k)}}$
35:                 **if** $R > \theta_{j,k}$ **then**
36:                     $SC \leftarrow (s_i^j, c_k)$
37:                 **end if**
38:             **end for**
39:         **end if**
40:         **end if**
41:     **end if**
42: **end if**

# 4 Proposed framework

## 4.1 Experimental setup

The proposed Distributed Load Balancing Mechanism for Controllers (DLBMT) has been evaluated using the NS2 framework, demonstrating a high level of authenticity. Within the DLBMT scheme, we selected RYU [16] as the experimental controller. RYU is an OpenFlow controller that is implemented in the Python programming language. The simulation experiments were carried out on a PC equipped with an Intel Core i7 CPU and 12GB of RAM. Additionally, the study calculated the average resource demand of PACKET-IN messages using the Iperf tool [17].

The performance of the proposed DLBMT mechanism was evaluated by applying it to modern network topologies in real-time scenarios, including Interroute and ARN, sourced from the Topology-zoo website. Furthermore, we incorporated the Atlanta and Germany50 topologies from the SND-lib, each with varying scales, to assess the effectiveness of the MTDLB methodology. A detailed summary of the network topology characteristics is provided in Table III.

Table 3: Indicated The Network Topologies Characteristics.

| topology | nodes | edges | controller | Capacity of controller(unit) |
|---|---|---|---|---|
| Atlanta | 15 | 22 | 3 | 2k |
| ARN | 30 | 29 | 4 | 2.5k |
| Germany50 | 50 | 88 | 5 | 3k |
| Interroute | 110 | 159 | 7 | 4k |

To evaluate the DLBMT performance, we compare it with four different schemes: DSMLB[11], SMSC[13], DLBM [14], DHA [15].

## 4.2 Simulation analysis

### 4.2.1 Performance evaluation of average response time

The response time criterion is utilized to evaluate the performance of the controller, as an increase in response time indicates an unbalanced load on the controller. As illustrated in Figure 4, the proposed DLBMT algorithm demonstrates a shorter response time compared to the DSMLB, DHA, DLBM, and SMSC algorithms across various topologies, including Atlanta, ARN, Germany50, and Interroute. The response time of the DLBMT mechanism has improved by 18%, 36%, 43%, and 53% compared to DSMLB, DHA, DLBM, and SMSC, respectively.

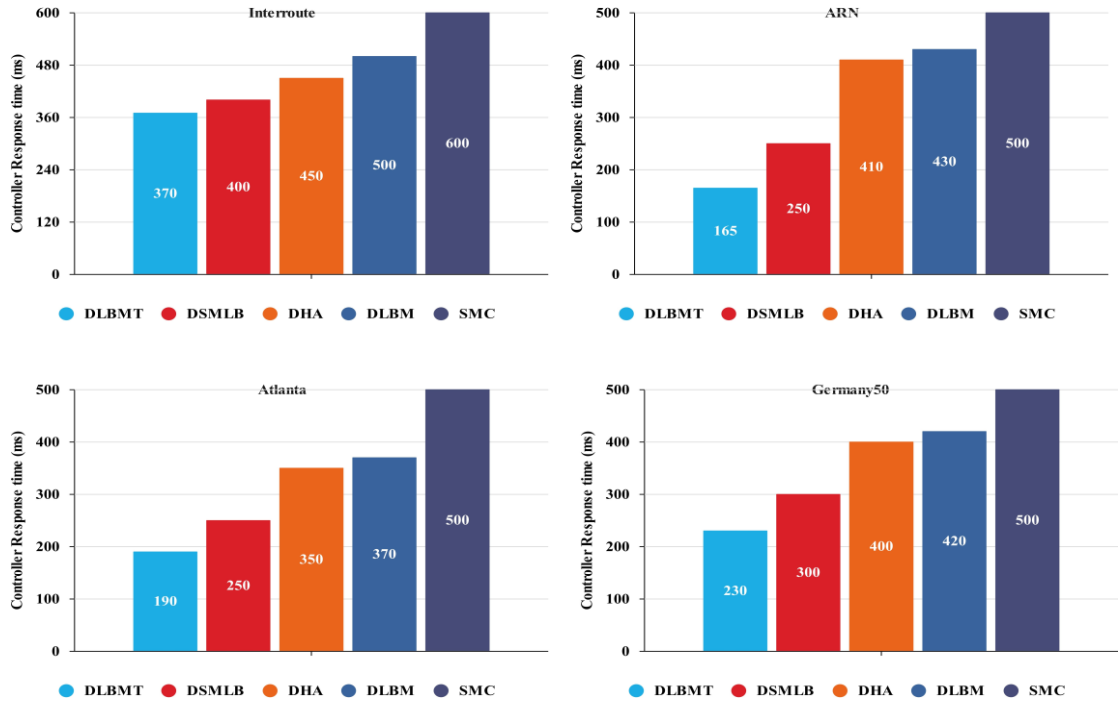

Figure 4: Performance Evaluation of Average Controller Response Time in topology Atlanta, Interroute, Germany50, and ARN.

### 4.2.2 Performance evaluation of migration cost

The migration cost is defined as the total resource expenditure incurred during the transfer of a switch, which includes the time taken for the migration, the bandwidth utilized, and any degradation in network performance resulting from the migration process. Migration cost in controller load balancing refers to the cost associated with moving a switch from one controller to another. This cost can include several factors, such as the time required to migrate the switch, the bandwidth consumed during the migration, and the impact on network performance during the migration. Minimizing migration cost is crucial in controller load balancing because frequent migrations can increase network overhead and reduce performance. To minimize migration costs, load balancing algorithms typically consider factors such as the current load on each controller, the available resources on each controller, and the migration cost associated with moving a switch from one controller to another.

By carefully balancing the load among controllers and minimizing migration costs, we can ensure that the network operates efficiently and effectively, while also minimizing the impact on network performance and reducing energy consumption. Yes, as shown in Figure 5, the migration cost of the proposed strategy is lower than the other four strategies in all topologies. This indicates that the proposed strategy is effective in selecting the appropriate switch and controller for migration, resulting in lower migration costs and improved network performance.

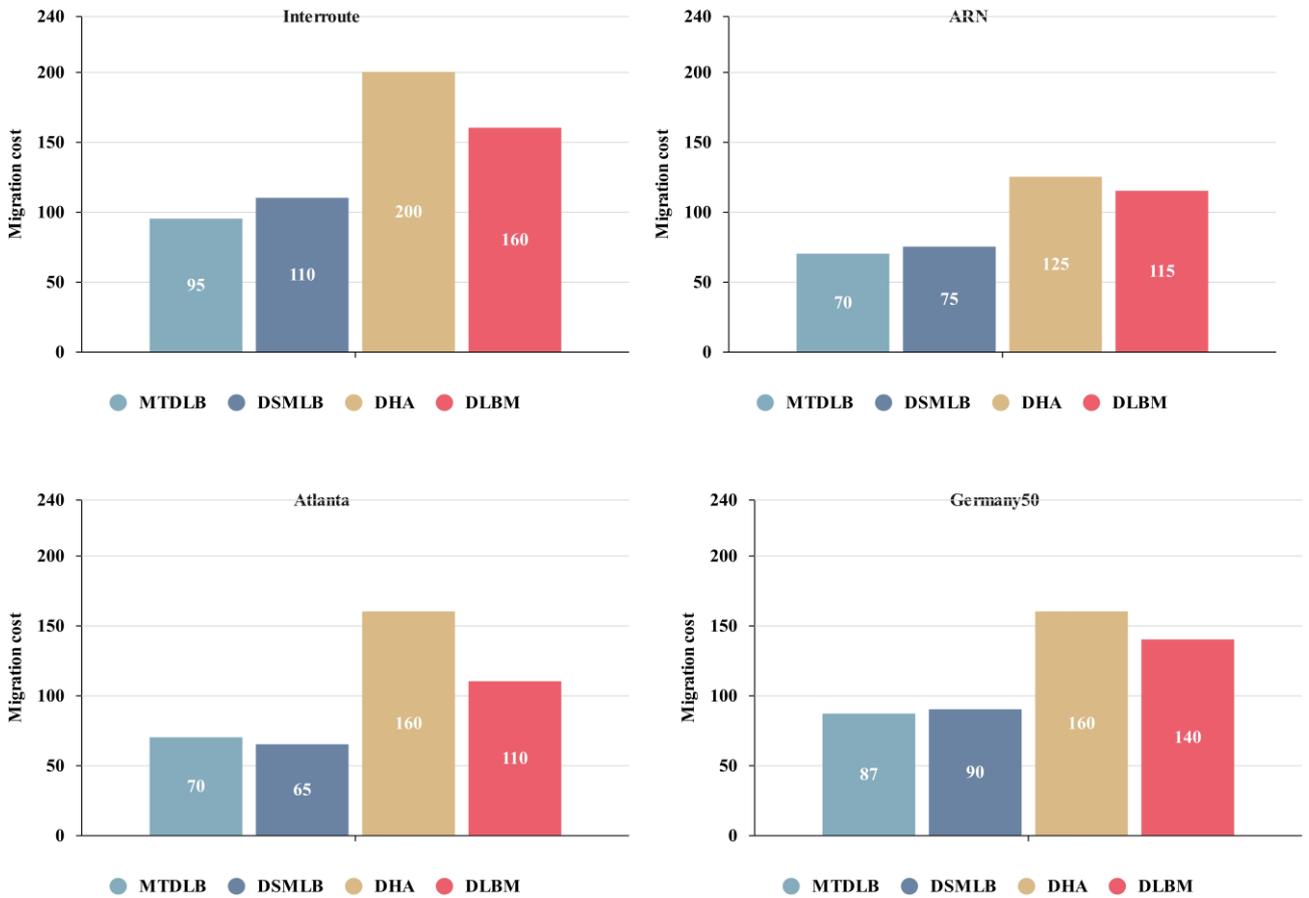

Figure 5: Performance Evaluation of migration cost in topology Interroute , Germany50 , Atlanta , and ARN.

### 4.2.3 Degree of load imbalance performance evaluation

The degree of load imbalance between controllers in controller load balancing refers to the difference in resource consumption or workload between controllers in a network. Load-balancing algorithms aim to distribute the workload or resource consumption evenly among the available controllers to improve network performance and efficiency. A high degree of load imbalance can lead to longer response times, decreased network availability, and higher energy consumption. By monitoring the degree of load imbalance and implementing appropriate load-balancing techniques, network administrators can ensure that the workload is distributed evenly among the available controllers, leading to improved network performance and reduced energy consumption.

In this section, the degree of imbalance in load strategy DLBMT and four other strategies (DSMLB, DHA, DLBM, SMSC) are compared in four different topologies. The proposed strategy has the least degree of imbalance in load in all topologies and under increasing transmission rates of PACKET-IN messages. As shown in Figure 6, the degree of load imbalance of the DLBMT strategy is lower than that of the other strategies under varying network loads.

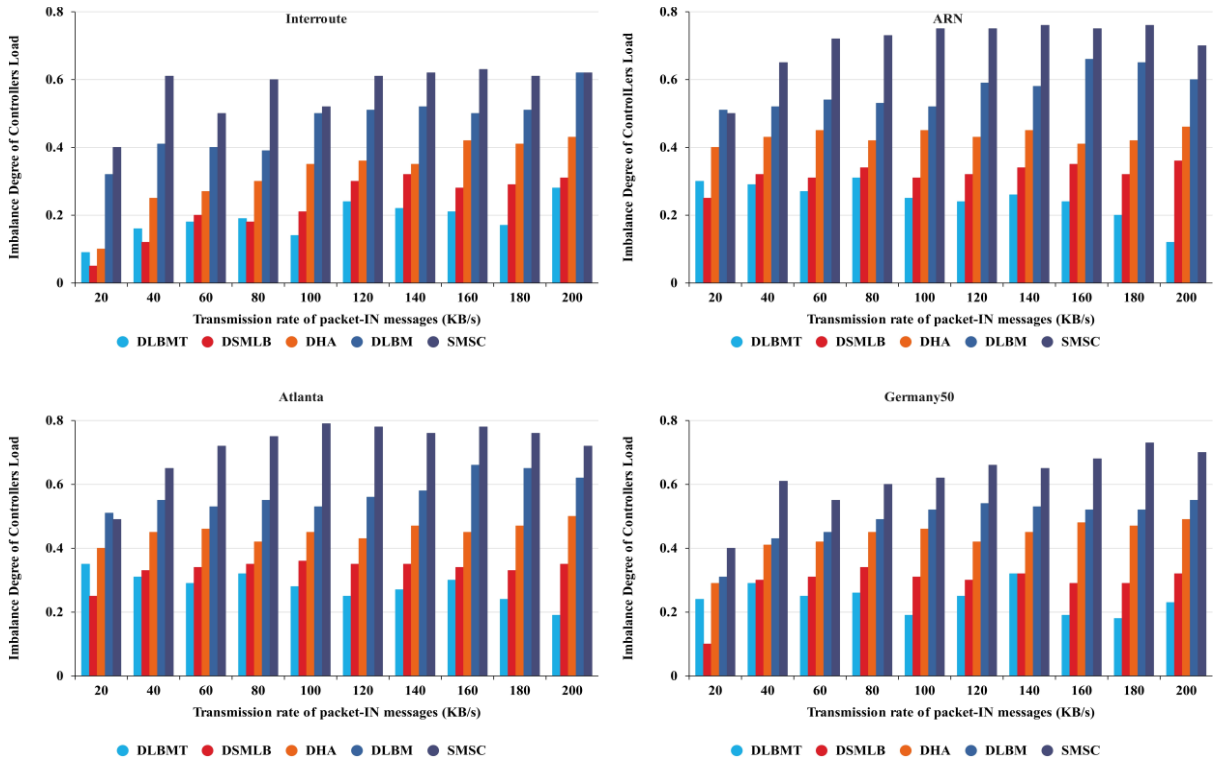

Figure 6: Performance Comparison on Degree of Load Imbalance in topology Interroute, Germany50, ARN, and Atlanta.

### 4.2.4 The Control plane load balancing rate

The control plane load balancing rate is a critical metric in software-defined networking (SDN) that measures the effectiveness and efficiency of distributing control loads across multiple controllers. This rate indicates how quickly and effectively the load is balanced across the control plane, ensuring that no single controller becomes overwhelmed while others remain underutilized. The importance of the control plane load balancing rate can be summarized as follows: A higher load balancing rate can enhance overall network performance by reducing response times and minimizing the likelihood of bottlenecks. Additionally, effective load balancing contributes to higher network reliability; should a controller fail or become overloaded, the system can swiftly redistribute the load to other controllers, thereby maintaining continuous network operation. Furthermore, efficient load balancing optimizes resource utilization across controllers, leading to lower operational costs and improved energy efficiency. Several factors influence the control plane load balancing rate.

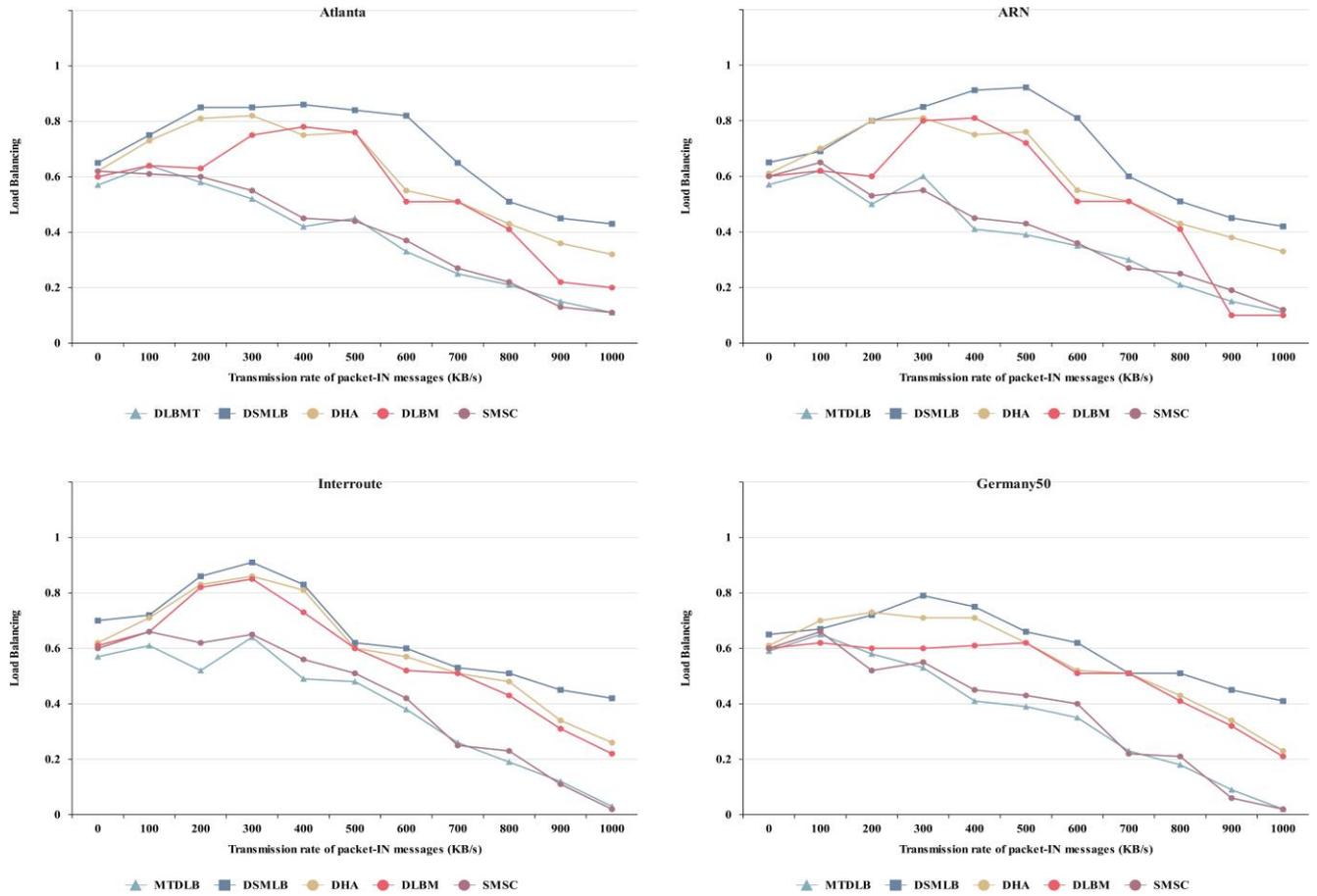

Figure 7: The control plane load balancing rate in topology Interroute, Germany50, ARN, and Atlanta.

### 4.2.5 Performance evaluation of communication overhead

In controller load balancing, communication overhead refers to the extra data and traffic that is generated by the controllers as they communicate with each other to balance the load. This communication overhead can impact the performance of the network, as it consumes bandwidth and processing resources that could otherwise be used for data transmission.

The proposed strategy results in less communication overhead between controllers and between controller-switch pairs compared to other strategies. The controller-to-controller communication overhead is very low due to multiple thresholds and sending messages to other controllers in case of a controller level change.

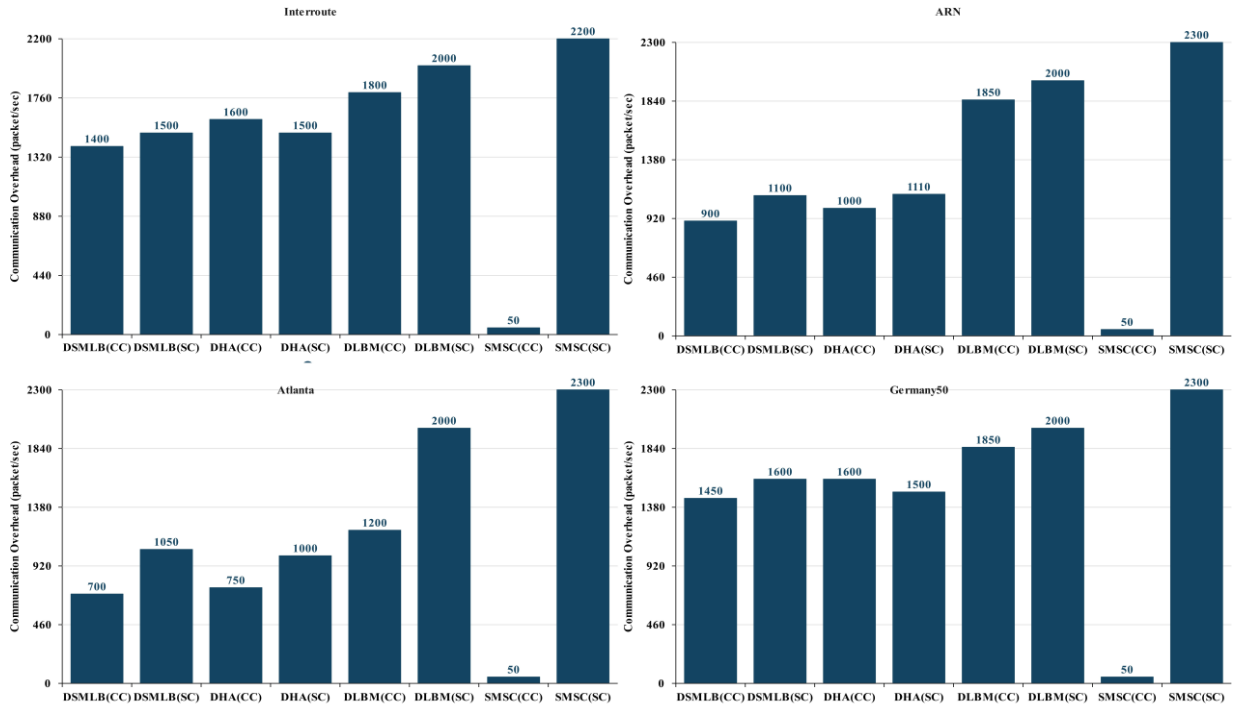

Figure 8: The performance evalution of communication overhead in topology Interroute, Germany50, ARN , and Atlanta.

### 4.2.6 Comparison Communication Delay, Reliability, and CPU and memory Consumption Metrics

In software-defined networking (SDN), it's crucial to evaluate various metrics such as communication delay, reliability, CPU consumption, and memory consumption to ensure optimal network performance. This section provides a comparative analysis of these metrics in different SDN strategies or configurations.

- **Communication Delay:** Communication delay refers to the time taken for a message to travel from one point to another in the network, encompassing transmission time, processing time, and queuing time. Factors influencing communication delay include network topology, which can increase delays due to longer paths and more hops, as well as the load balancing strategy, with different strategies potentially requiring varying levels of communication between controllers and switches.

- **Reliability:** Reliability is a crucial metric that measures the ability of the network to maintain consistent performance and availability, particularly in the face of failures or increased loads. Several factors influence reliability. One key factor is redundancy; the presence of backup controllers can significantly enhance the network's reliability by providing alternative options in case of a controller failure. Additionally, the efficiency of load balancing plays a vital role in ensuring reliability. Efficient load balancing helps prevent overloads, which can lead to system failures, thereby contributing to a more stable and dependable network operation.

- **CPU Consumption:** CPU consumption is defined as the amount of processing power utilized by the controllers and switches to manage network operations. Several factors influence CPU consumption in a network environment. One significant factor is the complexity of algorithms used for load balancing and routing; more complex algorithms typically consume more CPU resources as they require additional processing power to execute efficiently. Additionally, traffic volume plays a crucial role in determining CPU consumption; as traffic volumes increase, the need for processing also rises to maintain optimal performance levels across the network.

- **Memory Consumption:** Memory consumption measures the amount of memory used by network devices to store routing tables, flow entries, and other necessary data.

Table 4: Comparison Table of Communication Delay, Reliability, and Processor and Memory Consumption Metrics in Topology Interroute.

| Metrics | SMSC | DLBM | DHA | DSMLB | DLBMT |
|---|---|---|---|---|---|
| Communication Delay(ms) | 11.2 | 10.3 | 8.5 | 9.4 | 6.7 |
| Reliability(%) | 92.4% | 94.8% | 96.1% | 97.5% | 99.1% |
| CPU usage(%) | 49.8% | 47.3% | 41.8% | 44.2 % | 39.6% |
| Memory usage(MB) | 165.3 | 154.9 | 139.6 | 148.2 | 128.7 |

Table 5: Comparison Table of Communication Delay, Reliability, and Processor and Memory Consumption Metrics in Topology GERMANY50.

| Metrics | SMSC | DLBM | DHA | DSMLB | DLBMT |
|---|---|---|---|---|---|
| Communication Delay(ms) | 10.1 | 9.2 | 7.4 | 8.1 | 5.8 |
| Reliability(%) | 92.1% | 94.4% | 95.8 % | 97.2 % | 99.0% |
| CPU usage(%) | 46.7% | 44.0% | 39.4% | 41.7 % | 37.1% |
| Memory usage(MB) | 154.7 | 145.2 | 130.5 | 138.6 | 119.4 |

Table 6: Comparison Table of Communication Delay, Reliability, and Processor and Memory Consumption Metrics in Topology ARN.

| Metrics | SMSC | DLBM | DHA | DSMLB | DLBMT |
|---|---|---|---|---|---|
| Communication Delay(ms) | 9 | 8.3 | 6.6 | 7.3 | 5 |
| Reliability(%) | 91.8 % | 94.2% | 95.5 % | 96.8 % | 98.7% |
| CPU usage(%) | 45.2% | 42.9% | 37.6% | 40.2 % | 35.0% |
| Memory usage(MB) | 146 | 136.9 | 124.8 | 132.4 | 113.5 |

Table 7: Comparison Table of Communication Delay, Reliability, and Processor and Memory Consumption Metrics in Topology ATLANTA.

| Metrics | SMSC | DLBM | DHA | DSMLB | DLBMT |
|---|---|---|---|---|---|
| Communication Delay(ms) | 8 | 7.1 | 5.5 | 6.2 | 4.3 |
| Reliability(%) | 91.2% | 94.0% | 95.4 % | 98.5 % | 99.0% |
| CPU usage(%) | 44.1% | 41.5% | 36.4% | 38.9 % | 33.8% |
| Memory usage(MB) | 142.7 | 132.8 | 120.6 | 128.3 | 108.5 |

# 5   Conclusion:

This study aimed to address the challenges associated with controller load in software-defined networks (SDNs) by proposing a multi-level threshold approach for load balancing among controllers. The findings demonstrate that the proposed method effectively measures controller load using three critical metrics: CPU utilization, memory consumption, and bandwidth usage. By accurately assessing the status of controllers and strategically executing switch migration operations, the proposed framework significantly reduces response times across various traffic conditions compared to existing methods. Despite these promising results, several challenges remain that warrant further exploration. Notably, the diversity in controller capacities can affect the accuracy of load distribution, suggesting the need for more sophisticated algorithms capable of evaluating real-time controller capabilities. Additionally, the costs associated with switch migration operations pose a significant consideration; hence, future work should focus on optimizing these processes to minimize overhead while maximizing network efficiency.

Furthermore, the dynamic nature of network traffic necessitates the development of intelligent algorithms that can adaptively manage and distribute loads in real time. Incorporating machine learning techniques for traffic prediction and adjustment of load distribution strategies could be a valuable avenue for future research.In summary, this study lays the groundwork for enhancing load balancing in SDNs through innovative methodologies. Continued research in the areas of scalability, cost optimization, and adaptive load management will be essential for advancing the effectiveness of SDN architectures, ultimately leading to improved network performance and reliability.